\documentclass[ALICE,manyauthors]{cernphprep}

\usepackage[comma,square,numbers,sort&compress]{natbib}
\usepackage{hyperref}
\usepackage{lineno}
\usepackage{xcolor}
\usepackage{textcomp}
\usepackage[T1]{fontenc}
\usepackage{orcidlink}
\usepackage{xspace}

\usepackage[capitalize]{cleveref}

\newcommand{\rhopd}           {\ensuremath{\rho_{\mathrm{\overline{p}\overline{d}}}\xspace}}
\usepackage{xspace}

\newcommand{\PbPb}         {\mbox{Pb--Pb}\xspace}

\newcommand{\snn}          {\ensuremath{\sqrt{s_{\mathrm{NN}}}}\xspace}
\newcommand{\pt}           {\ensuremath{p_{\rm T}}\xspace}

\newcommand{\dEdx}         {\ensuremath{\textrm{d}E/\textrm{d}x}\xspace}

\newcommand{\nineH}        {$\sqrt{s}~=~0.9$~Te\kern-.1emV\xspace}
\newcommand{\seven}        {$\sqrt{s}~=~7$~Te\kern-.1emV\xspace}
\newcommand{\twoH}         {$\sqrt{s}~=~0.2$~Te\kern-.1emV\xspace}
\newcommand{\twosevensix}  {$\sqrt{s}~=~2.76$~Te\kern-.1emV\xspace}
\newcommand{\five}         {$\sqrt{s}~=~5.02$~Te\kern-.1emV\xspace}
\newcommand{\twosevensixnn}{$\sqrt{s_{\mathrm{NN}}}~=~2.76$~Te\kern-.1emV\xspace}
\newcommand{\fivenn}       {$\sqrt{s_{\mathrm{NN}}}~=~5.02$~Te\kern-.1emV\xspace}

\newcommand{\GeVc}         {Ge\kern-.1emV/$c$\xspace}
\newcommand{\MeVc}         {Me\kern-.1emV/$c$\xspace}
\newcommand{\TeV}          {Te\kern-.1emV\xspace}
\newcommand{\GeV}          {Ge\kern-.1emV\xspace}
\newcommand{\MeV}          {Me\kern-.1emV\xspace}
\newcommand{\GeVmass}      {Ge\kern-.2emV/$c^2$\xspace}
\newcommand{\MeVmass}      {Me\kern-.2emV/$c^2$\xspace}

\newcommand{\pbar}         {\ensuremath{\rm\overline{p}}\xspace}

\newcommand{\lmb}          {\ensuremath{\Lambda}\xspace}
\newcommand{\almb}         {\ensuremath{\overline{\Lambda}}\xspace}

\newcommand{\X}            {\ensuremath{\Xi^-}\xspace}
\newcommand{\Ix}           {\ensuremath{\overline{\Xi}^+}\xspace}

\newcommand{\dbar}   {\ensuremath{\overline{\mathrm{d}}}\xspace}
\newcommand{\dd}     {\ensuremath{\mathrm{d}}\xspace}

\newcommand{\dedx}{\mbox{\ensuremath{\dd E/\dd x}}\xspace}
\newcommand{\dvdy}{\mbox{\ensuremath{\dd V/\dd y}}\xspace}

\newcommand{\rhodL} {\ensuremath{\rho_{\bar{\mathrm{d}}\Delta\Lambda}}\xspace}
\newcommand{\kratioL} {\ensuremath{\kappa_{2}(\lmb - \almb)/\langle \lmb + \almb \rangle}\xspace}
\newcommand{\rhodp} {\ensuremath{\rho_{\dbar\pbar}}\xspace}
\newcommand{\rhokx} {\ensuremath{\rho_{\Delta\Xi\Delta\mathrm{K}}}\xspace}
\newcommand{\kratioP} {\ensuremath{\kappa_{2}(\rm p - \pbar)/\langle \rm p + \pbar \rangle}\xspace}
\newcommand{\kratioX} {\ensuremath{\kappa_{2}(\Ix - \X)/\langle \Ix + \X \rangle}}

\begin{document}%

\begin{titlepage}
\PHyear{2026}
\PHnumber{283}      
\PHdate{9 October}  
%

\title{Measurement of antideuteron and net-$\Lambda$ number fluctuations at the Large Hadron Collider}
\ShortTitle{Antideuteron and net-$\Lambda$ number fluctuations at the LHC}   
\hyphenation{nu-cleo-syn-the-sis}
\Collaboration{ALICE Collaboration}
\ShortAuthor{ALICE Collaboration} 

\begin{abstract}
The recent measurements of event-by-event correlations between the antideuteron and antiproton numbers in Pb--Pb collisions at the LHC deviate from the statistical hadronization model parameterizations that describe the yields and fluctuations of light-flavored hadrons. This discrepancy may stem from nucleosynthesis processes involved in the formation of composite states like antideuterons. In this letter, we present the first measurement of the antideuteron-net-$\Lambda$ number correlation obtained in Pb--Pb collisions with the ALICE detector.
This quantity is mostly sensitive to the quantum-number correlations induced by the hadronization process, as $\Lambda$ hyperons are not constituents of antideuterons.
Both the antideuteron--net-$\Lambda$ number correlation and the normalized second-order cumulant of the net-$\Lambda$ number are well described by the statistical hadronization model implemented in Thermal-FIST. This establishes strong evidence that the antideuteron-antiproton correlation is driven by the nucleosynthesis processes, while the correlation between antideuterons and non-nucleonic hadrons is primarily sensitive to the hadronization mechanism.
\end{abstract}
\end{titlepage}
\setcounter{page}{2}

The production of nuclei and antinuclei in heavy-ion collisions has been extensively studied in the last two decades~\cite{STAR:2011eej, ALICE:2015wav, STAR:2019sjh, ALICE:2022veq}.
Within the context of a multi-stage evolution of the system formed in such collisions, where particle abundances are fixed at the freeze-out of inelastic interactions, the measured abundances, of these states challenge our understanding of particle-production mechanisms: the nuclei produced, which have binding energies of $O(10\ \mathrm{MeV})$, must survive the extreme temperature, $O(100\ \mathrm{MeV})$, of the system created in the collision.
Strikingly, the average deuteron yield is described by both the coalescence model and the statistical hadronization model (SHM) across different colliding systems, whereas the production of other clusters, such as the hypertriton, is reproduced only by the nuclear coalescence model~\cite{ALICE:2022veq, ALICE:2024koa}.

The SHM describes the system as a hadron-resonance gas in full equilibrium at hadron emission, hence it predicts particle yields using as model parameters the volume ($V$), the baryon chemical potential ($\mu_B$), determining the net-baryon content of the system, and the temperature of the system at chemical freeze-out ($T_{\mathrm{chem}}$). The Grand Canonical Ensemble formulation of the SHM fits the measured production yields of light hadrons and nuclei in central Pb--Pb collisions at a center-of-mass energy per nucleon pair ($\snn$) of $\sqrt{s_{\mathrm{NN}}}=$2.76 TeV with $T_{\mathrm{chem}}~=156.5\pm1.5$~MeV and $\mu_B\sim0$~\cite{Andronic:2017pug}. Similar results have also been obtained in Pb--Pb collisions at $\snn=5.02$~TeV~\cite{ALICE:2022wpn}. On the other hand, the coalescence model uses a different approach to explain the production of nuclei: namely, nuclear clusters are formed by pre-existing nucleons, which merge when they are close in phase space. The size of the nucleon-emitting source, accessible through the analysis of femtoscopic correlations~\cite{PHENIX:2004yan, STAR:2004qya, ALICE:2015hvw, ALICE:2020ibs, ALICE:2025wuy}, the momentum distribution of the nucleons, as well as the nuclear wave function, are inputs that determine the formation probability of bound states~\cite{Scheibl:1998tk, Sun:2018mqq, Blum:2019suo}.
While measurements of average yields can be used to differentiate between these two mechanisms in small collision systems, such as pp and p--Pb, the discrimination power between the yields computed via SHM and coalescence in heavy-ion collisions is limited by the similarity between the source and nucleus sizes. In smaller systems, where the source size can be smaller than the nuclear size, the overlap between the wavefunctions of the initial nucleons and of the nuclear cluster is reduced, leading to a suppression of the nuclear abundances relative to SHM expectations~\cite{Bellini:2018epz, Sun:2018mqq, Bellini:2020cbj}. On the contrary, both models provide a quantitative description of yields of (anti)deuterons measured in Pb--Pb collisions at the LHC.
Hence, in order to differentiate between these two models, it is necessary to introduce novel observables that capture fluctuations and correlations beyond the average yield values.

In a recent study~\cite{ALICE:2022xiu}, the ALICE Collaboration has shown that the negative correlation measured between the antideuteron number and the antiproton number determined on an event-by-event basis can be used to discriminate among different nuclei production models.
The Pearson correlation coefficient, $\rho_{\mathrm{ab}}$, between different identified particles a and b can be expressed in terms of first and second-order cumulants, $\kappa_{1\{\mathrm{a,b}\}}$ and $\kappa_{2\{\mathrm{a,b}\}}$, as:

\begin{linenomath}
\begin{equation}
\label{eq7}
\kappa_{1\mathrm{a}} ~=~ \langle n_{\mathrm{a}} \rangle,
\end{equation}
\begin{equation}
\label{eq8}
\kappa_{2\mathrm{a}} ~=~ \langle (n_{\mathrm{a}} - \langle n_{\mathrm{a}} \rangle)^{2} \rangle,
\end{equation}
\begin{equation}
\label{eq9}
\rho_{\mathrm{ab}} ~=~ \langle (n_{\mathrm{a}} - \langle n_{\mathrm{a}} \rangle)(n_{\mathrm{b}} - \langle n_{\mathrm{b}} \rangle) \rangle/\sqrt{\kappa_{2\mathrm{a}}\kappa_{2\mathrm{b}}},
\end{equation}
\end{linenomath}

where $n_{\{\mathrm{a},\mathrm{b}\}}$ and $\langle n_{\{\mathrm{a},\mathrm{b}\}} \rangle$ are the event-by-event particle number and the event-averaged particle number of species a and b, respectively. The $\kappa_{1{\mathrm{a}}}$ ($\kappa_{1{\mathrm{b}}}$) and $\kappa_{2\mathrm{a}}$ ($\kappa_{2\mathrm{b}}$) are the first and second-order cumulants of the multiplicity distribution of species a (b).
The quantity $\rho_{\mathrm{ab}}$ provides information about the correlations between the abundances, and by extension the quantum numbers, of species a and b. In the system created in heavy-ion collisions, where quantum numbers are exactly conserved, the strength and sign of $\rho_{\mathrm{ab}}$ is driven by the correlations arising from quantum-number conservation. In particular, baryon–baryon pairs exhibit negative correlations as a consequence of baryon number conservation. Conversely, baryon-antibaryon pairs yield positive correlations.
When correlating nuclei with constituent baryons, additional correlations affect $\rho_{\mathrm{ab}}$ if the coalescence picture holds. Namely, the localized merging of nucleons affects the correlation range of baryon number~\cite{Feckova:2016kjx}. On the contrary, in a pure canonical ensemble (CE) SHM scenario, the baryon-number conservation is insensitive to the species correlated with the nucleus of interest: this is driven by the fraction of the total baryon number carried by each of the analyzed species within the experimental acceptance, rather than by a change in the baryon-number correlation range~\cite{Braun-Munzinger:2023gsd, Vovchenko:2020kwg}.
In Ref.~\cite{ALICE:2022xiu}, it was shown that an anticorrelation is qualitatively expected by both the SHM, using the CE to include the exact conservation of the baryon number, and coalescence models. However, neither model can quantitatively describe the data. Specifically, the CE SHM, implemented by the Thermal-FIST code~\cite{Vovchenko:2019pjl}, is able to reproduce the data only if the canonical volume $V_c$, over which quantum numbers are exactly conserved, is smaller than what is required for other light-flavor hadrons. Thus, event-by-event fluctuations of antideuterons represent an outlier from the perspective of the statistical hadronization picture valid for other species.

To assess the origin of this discrepancy, new observables are required, such as the Pearson correlation coefficient \rhodL between the antideuterons and net-$\Lambda$ number, i.e., $\Delta\Lambda=\Lambda - \bar{\Lambda}$. This quantity is primarily sensitive to the hadronization mechanism underlying the nuclear formation process, as it probes the correlation of the quantum numbers carried by a nucleus and a non-nucleonic hadron species whose production is not affected by the nucleosynthesis. As a cross check, the normalized second-order cumulant of the net-$\Lambda$ number, \kratioL, is also measured to determine whether $\Lambda$ fluctuations are consistently described by models that match the second-order fluctuations of other light-flavor hadrons, such as protons and charged $\Xi$ hyperons~\cite{ALICE:2022xpf, ALICE:2024rnr}, excluding light nuclei. This makes it possible to test for the independence of the properties of quantum-number conservation with respect to the quantum numbers carried by the species of interest.

In this Letter, we report the first measurement of both \kratioL and \rhodL in Pb--Pb collisions at the LHC. The results are extracted from the data collected by the ALICE Collaboration during the 2018 \PbPb run at \snn$~=~$5.02 TeV. In this Letter, we also improve on the precision of the published \rhodp results~\cite{ALICE:2022xiu} by combining the 2015 and the 2018 data samples. The increased sample size and the improved treatment of systematic uncertainties allow tight constraints to be imposed on nuclear production models in heavy-ion collisions. The analysis strategy is similar to that used in Ref.~\cite{ALICE:2022xiu}. Specifically, only antimatter (\dbar and \pbar) is measured to avoid the contamination coming from low-momentum secondary particles produced by the interaction of primary particles with the detector material. In the following, unless otherwise specified, the notation $\Lambda$ is used to denote both the hyperon and the antihyperon.

The ALICE detector and its performance are described in detail in Refs.~\cite{Aamodt:2008zz,Abelev:2014ffa}. Inelastic collision events are selected with a minimum bias (MB) trigger defined by a coincidence of signals in the V0 scintillator arrays~\cite{Abbas:2013taa}, located on both sides of the interaction point, covering the pseudorapidity, $\eta$, intervals $-3.7 <\eta <-1.7$ and $2.8<\eta<5.1$, respectively. An additional trigger based on the amplitude of the V0 signal is applied to enhance the selection of central (head-on) and semi-central collisions. Further selections are applied to reject events where more than one primary interaction vertex is reconstructed (pile-up)~\cite{Arslandok:2022dyb}. Finally, the position of the primary collision vertex along the beam direction is required to be within 8 cm from the nominal interaction point to benefit from the full acceptance of the ALICE central-barrel detectors. In total, about $9\times 10^7$ and $2.3\times 10^8$ events are selected from the 2015 and 2018 sample, respectively. The events are subdivided into eight centrality intervals defined by the V0 signal amplitude~\cite{Abelev:2013qoq, ALICE:2018tvk}.

The charged-particle tracks are reconstructed in the ALICE central barrel with the Inner Tracking System (ITS)~\cite{Abelev:2014ffa} and the Time Projection Chamber (TPC)~\cite{Alme:2010ke}, which are located within a solenoid that provides a homogeneous magnetic field of up to $0.5$ T in the direction of the beam axis. These two subsystems provide full azimuthal coverage for charged-particle trajectories in the pseudorapidity interval $|\eta| < 0.8$, which is used for all of the observables measured in this Letter.
The \dbar, \pbar, and \lmb candidates are identified in the transverse-momentum (\pt) ranges $0.6\leq p_{\mathrm{T}}^\mathrm{\overline{d}}<1.8\ \mathrm{GeV}/c$, $0.4\leq p_{\mathrm{T}}^{\mathrm{\overline{p}}}<0.9\ \mathrm{GeV}/c$, and $1.0\leq p_{\mathrm{T}}^{\Lambda}<4.0\ \mathrm{GeV}/c$, respectively, to select the bulk of the production with purity greater than $95\%$. For the \rhodp measurement, the \dbar acceptance is reduced to $0.8\leq p_{\mathrm{T}}<1.8\ \mathrm{GeV}/c$, so that the condition $p_{\mathrm{T}}^{\pbar}=p^{\dbar}_\mathrm{T}/2$ holds across the full data sample. This condition maximizes the sensitivity of \rhodp to nucleosynthesis effects, if present, according to momentum conservation in nuclear formation processes starting from pre-formed nucleons, enabling the separation between coalescence and statistical hadronization from the model-to-data comparison~\cite{ALICE:2022xiu}. While \dbar and \pbar are directly tracked in the detector, \lmb baryons are reconstructed via their weak decay to a proton and a charged pion, $\Lambda\to \mathrm{p} + \pi^-$. The charge-conjugated channel is used for the \almb reconstruction.

To guarantee a track-momentum resolution better than $2\%$ in the relevant $p_{\mathrm{T}}$ range and a specific energy loss (\dedx) resolution in the TPC of $5\%$, the selected tracks are required to have at least 70 out of a maximum possible 159 reconstructed space points in the TPC, and at least one hit in the two innermost layers of the ITS. This selection also ensures a resolution better than 300 $\mu$m~\cite{Abelev:2014ffa} on the distance of the closest approach to the primary vertex (DCA). In addition, the $\chi^{2}$ per space point in the TPC and the ITS from the track fit are required to be less than 2.5 and 36, respectively. For the \lmb decay products, the requirement of space points in the ITS is released to enable the reconstruction of the $\Lambda$ decay vertices, while a stricter threshold of 100 TPC space points is applied to reject candidates sharing common space points. Furthermore, tracks sharing more than 5 points in the TPC with other candidates are explicitly rejected. To avoid cross-contamination between the samples of \dbar and \lmb decay products, complementary DCA selections are applied to the two samples: $\vert\mathrm{DCA}^{\mathrm{\overline{d},\overline{p}}}\vert<0.1\ \mathrm{cm}$ and $\vert\mathrm{DCA}^{\Lambda\ \mathrm{decay\ prod.}}\vert>0.1\ \mathrm{cm}$, respectively.
The \lmb decay vertex is reconstructed using the same algorithm as in past measurements~\cite{ALICE:2013xmt, ALICE:2013wgn, ALICE:2016fzo}. The combinatorial background originating from the vertexing process is reduced by applying selections on the topological decay variables. They include the cosine of the pointing angle, $\cos\theta_{p}$, formed by the reconstructed hyperon momentum and the straight line connecting the production and decay vertices; the DCAs of the decay products to the primary vertex, as well as the DCA between the decay products themselves; the DCA of the \lmb candidate to the primary vertex. All events with at least two \lmb reconstructed from track pairs sharing at least one track are rejected to cancel spurious correlations.

The masses of the tracked particles are inferred by combining the measured momentum with the measurements of both \dEdx in the TPC gas and the flight time of the particle from the primary vertex to the Time-Of-Flight (TOF) detector~\cite{Akindinov:2013tea}. The particle identification (PID) selections in the TPC are expressed in terms of the resolution-normalized deviations from the expected PID response, computed with a Bethe--Bloch parameterization~\cite{Abelev:2014ffa} under the mass hypothesis of particle $i$, $n\sigma^{\rm TPC}_i$. The \dbar and \pbar candidates are identified with $|n\sigma^{\rm TPC}_{\rm d,p}|<2$, while $|n\sigma^{\rm TPC}_{i}|<4$ is required for the \lmb decay products. To reject the \pbar contamination in the \dbar sample for $p_{\mathrm{T}}>1\ \mathrm{GeV}/c$, the particle mass, determined from the flight time and integrated track length measurements, is required to be compatible with the nominal mass~\cite{ParticleDataGroup:2024cfk} within $2.5\sigma$, where $\sigma$ is the TOF mass resolution. The TOF measurement is also used to reject the positron contamination in the proton sample for $p_{\mathrm{T}}>0.6\ \mathrm{GeV}/c$. The \lmb candidates are selected within a $2\sigma$ invariant-mass window centered around the nominal \lmb mass~\cite{ParticleDataGroup:2024cfk}, $\sigma$ being the invariant-mass resolution in this case.

The cumulants computed from the number of candidates identified in each event are corrected for the reconstruction and selection efficiency estimated from Monte Carlo (MC) simulations based on the HIJING~\cite{Wang:1991hta} event generator and the GEANT4~\cite{GEANT4:2002zbu} transport code. To enhance the simulated antideuteron sample, 10 antideuterons are injected into each HIJING event with realistic $p_{\mathrm{T}}$ distributions. The reconstruction and selection efficiency of \dbar is about 70\% and 25\% in the TPC and TOF, respectively, while the \lmb efficiency ranges from 5\% to 25\%, increasing with increasing $p_{\mathrm{T}}$. The efficiency is centrality dependent, with a reduction by about 10\% in the most central collisions with respect to peripheral events due to the higher detector occupancy. The analytical expressions of the efficiency-corrected cumulants are derived under the hypothesis of a binomial response of the detectors~\cite{Nonaka:2017kko}, which has been verified with a dedicated MC study. The agreement with the binomial hypothesis is ensured by the aforementioned candidate selections. The centrality-binned efficiency-corrected cumulants are also affected by volume fluctuations (VF), which are related to the fluctuations in the number of participant nucleons within a fixed centrality interval~\cite{Braun-Munzinger:2016yjz}. At the LHC, where matter and antimatter are balanced~\cite{ALICE:2023ulv}, the VF cancel out for event-by-event observables involving net-particle cumulants up to the third order~\cite{Braun-Munzinger:2016yjz, Rustamov:2022sqm, Holzmann:2024wyd}: this is the case for \kratioL. On the other hand, the impact of VF on observables such as \rhodL and \rhodp must be explicitly assessed. In this work, VF are determined via the event-mixing described in Refs.~\cite{Rustamov:2022sqm, Holzmann:2024wyd}. In this approach, the mixed-event tracks are extracted from uncorrelated events, up to the track multiplicity of the initial event, thus preserving the original VF. To mitigate the impact of correlated sources, such as resonance decays into charged particles, the mixing is carried out using only same-sign tracks. Finally, the cumulants of the volume distributions are extracted from measurements of antiproton cumulants in the mixed-event sample up to the second order, assuming that antiprotons are fully uncorrelated in the mixed events. While \rhopd~shows a sizable increase in absolute value by a factor of about 3 due to VF, the correction has a negligible effect in \rhodL due to the suppression of VF in the $\Delta\Lambda$ cumulants~\cite{Braun-Munzinger:2016yjz}.
The full analysis procedure, including the applied corrections, has been validated through a MC closure check. This consists of comparing the generated MC observables with the results obtained by applying the full analysis chain to MC reconstructed data. In this work, an agreement within $1.5\sigma$ is observed for all of the analyzed observables, across the 0-80\% centrality range.

The statistical uncertainties on the corrected observables are obtained using the subensemble method~\cite{Anticic:2013htn, ALICE:2017jsh, Pandav:2018bdx}. The systematic uncertainties are extracted by varying the selections on the aforementioned track and vertex variables around their nominal values. The selections on the \pbar and \dbar PID variables and \lmb invariant-mass are varied as well. The analysis is repeated multiple times with different combinations of these criteria: in each centrality interval, the standard deviation of the distribution of the fully-corrected observables is assigned as a systematic uncertainty. The dominant systematic contribution is given by the variation of DCA and primary-vertex $z$ selections, which correspond to about $5\times10^{-4}$ in \rhodL and \rhodp, and about $5\times10^{-3}$ in \kratioL, respectively. The uncertainty contributions, both correlated and uncorrelated with centrality, are separately estimated in this procedure. The correlated component is determined by analyzing centrality-correlated shifts in the observables obtained with varied analysis selections. The correlated uncertainty is then deconvolved from the systematic uncertainty estimated in each of the centrality intervals separately. The correlated systematic uncertainty is about $3\times10^{-4}$ for \rhodp and \rhodL, and about $4\times10^{-3}$ for \kratioL.

\begin{figure}[h]
    \centering
    \includegraphics[width=\textwidth]{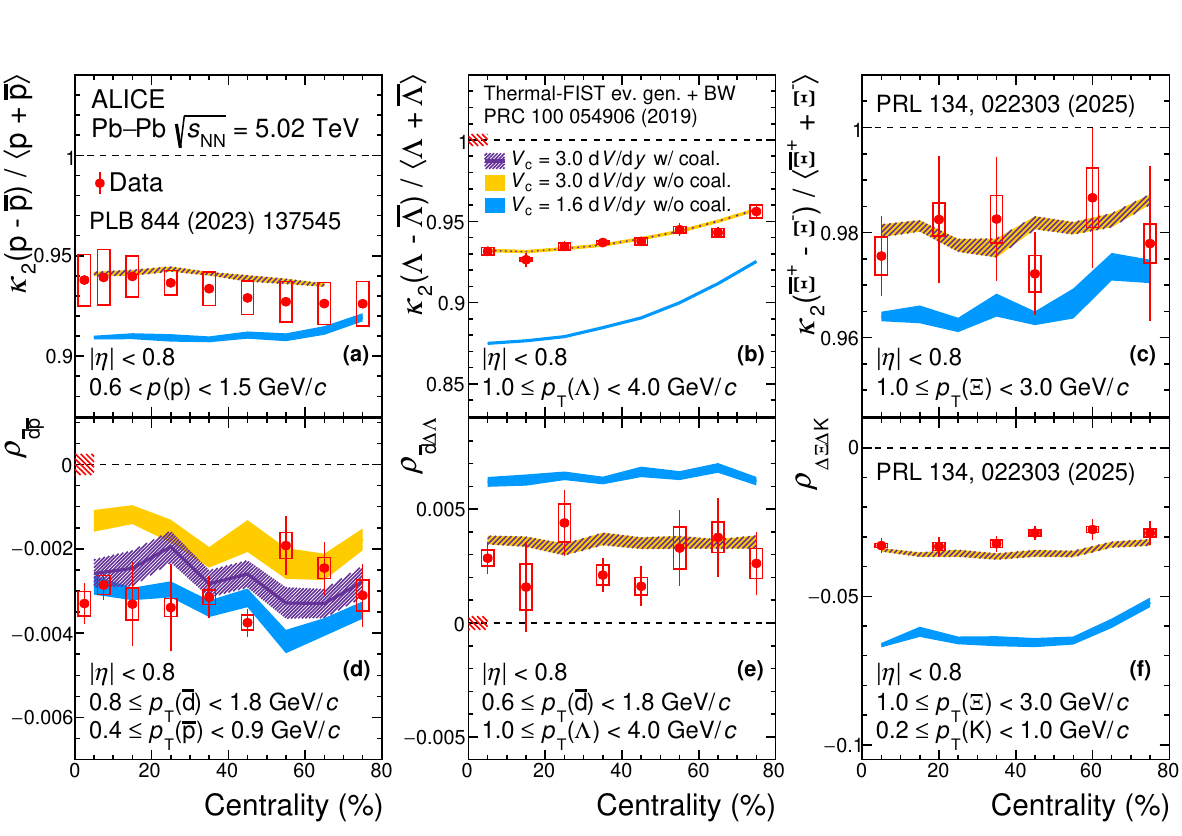}
    \caption{Upper panels: normalized second-order cumulants \kratioP~\cite{ALICE:2022xpf}, \kratioL, and \kratioX~\cite{ALICE:2024rnr}. Lower panels: Pearson correlation coefficients \rhodp~\cite{ALICE:2022xiu}, \rhodL, and \rhokx~\cite{ALICE:2024rnr}. The results of the analysis described in this Letter are shown in panels (b), (d), and (e): (b) and (e) are obtained from the 2018 data, while (d) is obtained with the combined 2015+2018 data. The observables are measured as a function of centrality in Pb--Pb collisions at $\sqrt{s_{\mathrm{NN}}}=5.02\ \mathrm{TeV}$. The error bars and boxes show the statistical and systematic uncertainties, respectively. For \kratioL, \rhodp, and \rhodL, the boxes and shaded areas indicate the centrality uncorrelated and correlated components of the systematic uncertainty, respectively. The correlated components are centered around the value of the uncorrelated particle-production baseline, which is represented as a dashed black line in each panel.
    The results are compared with model calculations, shown as colored bands, obtained with the Thermal-FIST model~\cite{Vovchenko:2019pjl}, both with and without nuclear coalescence effects, and for different $V_c$ values. The yellow bands with purple shades correspond to the cases where the calculations with and without coalescence effects coincide. The width of the band represents the statistical uncertainty in the model calculations.}
    \label{fig:fluct}
\end{figure}

In \cref{fig:fluct}, the fully corrected \kratioL and \rhodL are shown as a function of centrality in panels (b) and (e), respectively, along with other event-by-event fluctuation observables measured by the ALICE Collaboration: namely the normalized second-order cumulants \kratioP~\cite{ALICE:2022xpf} and \kratioX~\cite{ALICE:2024rnr}, and the Pearson correlation coefficients \rhodp~\cite{ALICE:2022xiu} and \rhokx~\cite{ALICE:2024rnr}, shown in panels (a), (c), (d), and (f), respectively.
All of these observables are compared with the expected values for the uncorrelated particle-production, which is 0 for the Pearson correlation coefficients and 1 for the normalized second-order cumulants.
The results are compared with calculations obtained with the Thermal-FIST~\cite{Vovchenko:2019pjl} implementation of the statistical hadronization model in the canonical ensemble. In this model, the exact conservation of quantum numbers is required across a truncated fireball using a box approximation: the fireball size is controlled by the canonical volume parameter, $V_c$, expressed as multiples of the volume corresponding to one rapidity unit, \dvdy. The exact value of the $V_c$ parameter depends on the specific implementation of quantum-number conservation in the model. The thermal parameters of the models, $T$, \dvdy, and $\gamma_s$, were extracted in previous studies by comparing model calculations with the yield ratios of several hadron species to charged pions measured as a function of the average charged-particle multiplicity~\cite{Vovchenko:2019kes}. The strangeness-saturation parameter $\gamma_s$, which acts as a multiplicative factor in the CE SHM partition function to describe incomplete strangeness equilibration, is equal to 1 in head-on Pb--Pb collisions, and is less than 1 in peripheral Pb–Pb collisions (approximately 0.9) because of incomplete strangeness equilibration. The event generator implemented in Thermal-FIST is used, employing the $p_{\mathrm{T}}$ distributions parametrized using the Blast-Wave model tuned to combined fits of the measured light-flavor hadron spectra~\cite{ALICE:2019hno, ALICE:2020nkc}. The model calculations are reported for two different values of $V_c$. On the one hand, $V_c=3.0~\dvdy$, which is independently obtained from the model-to-data comparison of both the hadron-yield ratios and the fluctuation observables \kratioX and \rhokx~\cite{Vovchenko:2019kes, ALICE:2024rnr}; while on the other hand, $V_c=1.6~\dvdy$ is the best value to describe \rhodp~\cite{ALICE:2022xiu}. Thermal-FIST calculations are carried out for fixed system volumes, hence they do not require any additional VF corrections.

The normalized second-order cumulant \kratioL obtained in this Letter shows a significant deviation from the uncorrelated particle-production baseline across the analyzed centrality range. This result is interpreted as an effect of the exact strangeness and baryon number conservation in the fireball, in line with previous collider measurements~\cite{STAR:2020ddh}.
Additionally, the data points show an increasing trend going from central to peripheral collisions. This is driven by a decrease of the \lmb and \almb acceptance, caused by the interplay between the fixed \pt acceptance of the present measurement and the evolution of the $p_{\mathrm{T}}$ spectra of hadrons across centrality as a result of space-momentum correlations~\cite{ALICE:2013xmt, ALICE:2025cqy}. The experimental results are well described by the Thermal-FIST calculations based on $V_c=3.0~\dvdy$, within the truncated-fireball approximation, while the deviation from unity predicted by $V_c=1.6~\dvdy$ is larger than that measured. These results, which are in line with past (net)-baryon fluctuation measurements, suggesting that the exact conservation laws of baryon number and strangeness affect the light-flavor hadronization in a rapidity range that is consistent for all species, independently of the specific quantum numbers carried by the analyzed species.

The Pearson correlation coefficient \rhodL shows a positive correlation, which is driven by the conservation of baryon number in the underlying hadronization process.
The statistical significance of this correlation, corresponding to 3.4$\sigma$, is determined by evaluating the probability $P(\chi^2>\chi^2_{\mathrm{obs}})$, where the $\chi^2_{\mathrm{obs}}$ quantifying the model-to-data deviation is computed using the full covariance matrix of the measured points as a function of centrality. The data are also compared with the Thermal-FIST calculations based on different model hypotheses. The calculations based on $V_c=1.6~\dvdy$ show a clear deviation from the data. Its significance, determined with this procedure, is 6.6$\sigma$. In contrast, the model expectations for $V_c=3.0~\dvdy$ well describe the data within 0.8$\sigma$. This observation is consistent with the previous measurements involving strange and non-strange baryons. Hence, within the present model assumptions, these results suggest that, in Pb--Pb collisions, the conservation of baryon number is consistently regulated by a single hadronization mechanism, independently of the analyzed light-flavored hadron species. This provides the baseline upon which nucleosynthesis reactions occur.

The results for \rhodp are shown in the lower left panel of \cref{fig:fluct}. The data confirm the anticorrelation already observed in the previous study~\cite{ALICE:2022xiu}, now including a full-fledged data-driven correction of VF. In addition, the data show a strong deviation of 3.9$\sigma$ from the Thermal-FIST calculations based on $V_c=3.0~\dvdy$. As already discussed in Ref.~\cite{ALICE:2022xiu}, these data can be quantitatively described using $V_c=1.6~\dvdy$. This is verified also in the present work, where an agreement within 1.8$\sigma$ is obtained between model and data. This result implies that the baryon-number correlation between \dbar and \pbar, having an acceptance $p_{\mathrm{T}}^{\pbar}=p^{\dbar}_\mathrm{T}/2$ over the analyzed sample, is stronger than that observed for \rhodL and \kratioL, as well as for the other measurements reported in \cref{fig:fluct}. This further suggests that the correlations between \dbar and its constituents are formed later than the early correlations tested with the other hadrons. In terms of the model parameters, larger $V_c$ values are qualitatively connected to longer rapidity correlation ranges, which form earlier in the system evolution. This indicates that the antideuteron-antiproton correlation might be sensitive to nuclear formation mechanisms, such as nuclear coalescence, acting on top of the nucleon production process. To this end, a calculation was carried out using a realistic coalescence afterburner based on the Wigner formalism~\cite{Mahlein:2023fmx, Mahlein:2024pur}. In this model, the nuclear coalescence is applied on an event-by-event basis to Pb--Pb events produced with the Thermal-FIST event generator, using $V_c=3.0~\dvdy$ and excluding (anti)deuterons from the list of generated hadrons. Strong and electromagnetic decays of resonances are also included in the model. The nucleon-emitting source size parametrization is obtained from the p-p femtoscopic correlations measured in Pb--Pb collisions at the LHC~\cite{ALICE:2025wuy}, while the Wigner function is computed using the Argonne $v_{18}$ wave function for the antideuteron~\cite{Wiringa:1994wb, Mahlein:2023fmx, Mahlein:2024pur}.
The calculations based on this model are shown in \cref{fig:fluct}. The coalescence-driven expectations differ from the Thermal-FIST baseline of $V_c=3.0~\dvdy$ only for \rhodp, while all other observables reported in \cref{fig:fluct} are unaffected by the nuclear formation mechanism, as they do not correlate nuclei with nucleons. In addition, it is observed that the nuclear coalescence increases the correlation strength for \rhodp compared to the full statistical hadronization scenario using $V_c=3.0~\dvdy$, reaching values that can only be obtained with a smaller $V_{c}$ without invoking any additional (anti)nucleosynthesis mechanism. The coalescence-based calculations agree with the data within 1.5$\sigma$. These considerations support the hypothesis that in heavy-ion collisions (anti)deuterons are formed by the fusion of (anti)nucleons produced by a locally equilibrated fireball.

In summary, the results reported in this Letter indicate that the event-by-event production of antideuterons is a sensitive probe of the nuclear formation process. The measured \rhodL suggests that the hadronization mechanism for the nucleons composing the deuterons is consistent with the universal picture valid also for other light-flavor species, as quantified by the similar canonical volume parameter used in Thermal-FIST. On the other hand, \rhodp is affected by the nucleosynthesis process, which induces additional local baryon-number correlations. By taking into account the nucleosynthesis process via a coalescence approach based on the Wigner function formalism, it is shown that also the \rhodp fits in the same hadronization description valid for the light-flavored species, indicating that deuterons are formed through the coalescence of the nucleons produced in statistical equilibrium. On the contrary, it is not possible to describe \rhodp in a baryon-number conservation picture consistent for all species, here represented by Thermal-FIST with $V_c=3.0~\dvdy$, unless nuclear coalescence is invoked. While the specific $V_c$ value depends on the model implementation, its consistency across different species suggests a uniform hadronization picture across different hadrons including nuclear coalescence. In the future, the box-approximation treatment of the conservation of multiple quantum numbers in the Thermal-FIST model could be improved leveraging recent modeling efforts to provide a more complete description of local rapidity correlations, as well as enabling the measurement of the baryon-number and strangeness rapidity correlation length~\cite{Braun-Munzinger:2023gsd, Vovchenko:2024pvk}. In addition to this, an improved tuning of the coalescence afterburner could be achieved by exploiting the measured (anti)deuteron yields as a function of centrality in Pb--Pb collisions.
On the experimental side,
the measurement of the Pearson correlation coefficient could be extended both to $A=3$ (hyper)nuclei and down to p--Pb and pp collision systems taking advantage of the higher luminosities available in the ongoing Run 3 and upcoming Run 4 of the LHC. These measurements will be crucial to fully comprehend the formation of light nuclei in the laboratory.

\newenvironment{acknowledgement}{\relax}{\relax}
\begin{acknowledgement}
\section*{Acknowledgements}

The ALICE Collaboration would like to thank all its engineers and technicians for their invaluable contributions to the construction of the experiment and the CERN accelerator teams for the outstanding performance of the LHC complex.
The ALICE Collaboration gratefully acknowledges the resources and support provided by all Grid centres and the Worldwide LHC Computing Grid (WLCG) collaboration.
The ALICE Collaboration acknowledges the following funding agencies for their support in building and running the ALICE detector:
A. I. Alikhanyan National Science Laboratory (Yerevan Physics Institute) Foundation (ANSL), State Committee of Science and World Federation of Scientists (WFS), Armenia;
Austrian Academy of Sciences, Austrian Science Fund (FWF): [M 2467-N36] and Nationalstiftung f\"{u}r Forschung, Technologie und Entwicklung, Austria;
Ministry of Communications and High Technologies, National Nuclear Research Center, Azerbaijan;
Rede Nacional de Física de Altas Energias (Renafae), Financiadora de Estudos e Projetos (Finep), Funda\c{c}\~{a}o de Amparo \`{a} Pesquisa do Estado de S\~{a}o Paulo (FAPESP) and The Sao Paulo Research Foundation  (FAPESP), Brazil;
Bulgarian Ministry of Education and Science, within the National Roadmap for Research Infrastructures 2020-2027 (object CERN), Bulgaria;
Ministry of Education of China (MOEC) , Ministry of Science \& Technology of China (MSTC) and National Natural Science Foundation of China (NSFC), China;
Ministry of Science and Education and Croatian Science Foundation, Croatia;
Centro de Aplicaciones Tecnol\'{o}gicas y Desarrollo Nuclear (CEADEN), Cubaenerg\'{\i}a, Cuba;
Ministry of Education, Youth and Sports of the Czech Republic, Czech Republic;
The Danish Council for Independent Research | Natural Sciences, the VILLUM FONDEN and Danish National Research Foundation (DNRF), Denmark;
Helsinki Institute of Physics (HIP), Finland;
Commissariat \`{a} l'Energie Atomique (CEA) and Institut National de Physique Nucl\'{e}aire et de Physique des Particules (IN2P3) and Centre National de la Recherche Scientifique (CNRS), France;
Bundesministerium f\"{u}r Forschung, Technologie und Raumfahrt (BMFTR) and GSI Helmholtzzentrum f\"{u}r Schwerionenforschung GmbH, Germany;
National Research, Development and Innovation Office, Hungary;
Department of Atomic Energy Government of India (DAE), Department of Science and Technology, Government of India (DST), University Grants Commission, Government of India (UGC) and Council of Scientific and Industrial Research (CSIR), India;
National Research and Innovation Agency - BRIN, Indonesia;
Istituto Nazionale di Fisica Nucleare (INFN), Italy;
Japanese Ministry of Education, Culture, Sports, Science and Technology (MEXT) and Japan Society for the Promotion of Science (JSPS) KAKENHI, Japan;
Consejo Nacional de Ciencia (CONACYT) y Tecnolog\'{i}a, through Fondo de Cooperaci\'{o}n Internacional en Ciencia y Tecnolog\'{i}a (FONCICYT) and Direcci\'{o}n General de Asuntos del Personal Academico (DGAPA), Mexico;
Nederlandse Organisatie voor Wetenschappelijk Onderzoek (NWO), Netherlands;
The Research Council of Norway, Norway;
Pontificia Universidad Cat\'{o}lica del Per\'{u}, Peru;
Ministry of Science and Higher Education, National Science Centre and WUT ID-UB, Poland;
Korea Institute of Science and Technology Information and National Research Foundation of Korea (NRF), Republic of Korea;
Ministry of Education and Scientific Research, Institute of Atomic Physics, Ministry of Research and Innovation and Institute of Atomic Physics and Universitatea Nationala de Stiinta si Tehnologie Politehnica Bucuresti, Romania;
Ministerstvo skolstva, vyskumu, vyvoja a mladeze SR, Slovakia;
National Research Foundation of South Africa, South Africa;
Swedish Research Council (VR) and Knut \& Alice Wallenberg Foundation (KAW), Sweden;
European Organization for Nuclear Research, Switzerland;
Suranaree University of Technology (SUT), National Science and Technology Development Agency (NSTDA) and National Science, Research and Innovation Fund (NSRF via PMU-B B05F650021), Thailand;
Turkish Energy, Nuclear and Mineral Research Agency (TENMAK), Turkey;
National Academy of  Sciences of Ukraine, Ukraine;
Science and Technology Facilities Council (STFC), United Kingdom;
National Science Foundation of the United States of America (NSF) and United States Department of Energy, Office of Nuclear Physics (DOE NP), United States of America.
In addition, individual groups or members have received support from:
Czech Science Foundation (grant no. 23-07499S), Czech Republic;
FORTE project, reg.\ no.\ CZ.02.01.01/00/22\_008/0004632, Czech Republic, co-funded by the European Union, Czech Republic;
European Research Council (grant no. 101220549), European Union;
Deutsche Forschungs Gemeinschaft (DFG, German Research Foundation) ``Neutrinos and Dark Matter in Astro- and Particle Physics'' (grant no. SFB 1258), Germany;
CONVECS project, CUP C97H23001700002 FESR 2021-2027 program, Italy.
\end{acknowledgement}

\bibliographystyle{utphys}   
\bibliography{biblio}

\providecommand{\href}[2]{#2}\begingroup\raggedright\begin{thebibliography}{10}

\bibitem{STAR:2011eej}
{\bfseries STAR} Collaboration, H.~Agakishiev {\em et~al.}, ``{Observation of
  the antimatter helium-4 nucleus}'',
  \href{https://doi.org/10.1038/nature10079}{{\em Nature} {\bfseries 473}
  (2011) 353}, \href{https://arxiv.org/abs/1103.3312}{{\ttfamily
  arXiv:1103.3312 [nucl-ex]}}. [Erratum: Nature 475, 412 (2011)].

\bibitem{ALICE:2015wav}
{\bfseries ALICE} Collaboration, J.~Adam {\em et~al.}, ``{Production of light
  nuclei and anti-nuclei in pp and Pb-Pb collisions at energies available at
  the CERN Large Hadron Collider}'',
  \href{https://doi.org/10.1103/PhysRevC.93.024917}{{\em Phys. Rev. C}
  {\bfseries 93} (2016) 024917},
  \href{https://arxiv.org/abs/1506.08951}{{\ttfamily arXiv:1506.08951
  [nucl-ex]}}.

\bibitem{STAR:2019sjh}
{\bfseries STAR} Collaboration, J.~Adam {\em et~al.}, ``{Beam energy dependence
  of (anti-)deuteron production in Au + Au collisions at the BNL Relativistic
  Heavy Ion Collider}'', \href{https://doi.org/10.1103/PhysRevC.99.064905}{{\em
  Phys. Rev. C} {\bfseries 99} (2019) 064905},
  \href{https://arxiv.org/abs/1903.11778}{{\ttfamily arXiv:1903.11778
  [nucl-ex]}}.

\bibitem{ALICE:2022veq}
{\bfseries ALICE} Collaboration, S.~Acharya {\em et~al.}, ``{Light (anti)nuclei
  production in Pb-Pb collisions at $\sqrt{s_{\rm NN}}$=5.02~TeV}'',
  \href{https://doi.org/10.1103/PhysRevC.107.064904}{{\em Phys. Rev. C}
  {\bfseries 107} (2023) 064904},
  \href{https://arxiv.org/abs/2211.14015}{{\ttfamily arXiv:2211.14015
  [nucl-ex]}}.

\bibitem{ALICE:2024koa}
{\bfseries ALICE} Collaboration, S.~Acharya {\em et~al.}, ``{Measurement of
  {\ensuremath{_\Lambda^{3}}}H production in Pb{\textendash}Pb collisions at
  $\sqrt{s_{\rm NN}}$=5.02 TeV}'',
  \href{https://doi.org/10.1016/j.physletb.2024.139066}{{\em Phys. Lett. B}
  {\bfseries 860} (2025) 139066},
  \href{https://arxiv.org/abs/2405.19839}{{\ttfamily arXiv:2405.19839
  [nucl-ex]}}.

\bibitem{Andronic:2017pug}
A.~Andronic, P.~Braun-Munzinger, K.~Redlich, and J.~Stachel, ``{Decoding the
  phase structure of QCD via particle production at high energy}'',
  \href{https://doi.org/10.1038/s41586-018-0491-6}{{\em Nature} {\bfseries 561}
  (2018) 321--330}, \href{https://arxiv.org/abs/1710.09425}{{\ttfamily
  arXiv:1710.09425 [nucl-th]}}.

\bibitem{ALICE:2022wpn}
{\bfseries ALICE} Collaboration, S.~Acharya {\em et~al.}, ``{The ALICE
  experiment: a journey through QCD}'',
  \href{https://doi.org/10.1140/epjc/s10052-024-12935-y}{{\em Eur. Phys. J. C}
  {\bfseries 84} (2024) 813},
  \href{https://arxiv.org/abs/2211.04384}{{\ttfamily arXiv:2211.04384
  [nucl-ex]}}.

\bibitem{PHENIX:2004yan}
{\bfseries PHENIX} Collaboration, S.~S. Adler {\em et~al.}, ``{Bose-Einstein
  correlations of charged pion pairs in Au + Au collisions at
  $\sqrt{s_{\mathrm{NN}}}$ = 200 GeV}'',
  \href{https://doi.org/10.1103/PhysRevLett.93.152302}{{\em Phys. Rev. Lett.}
  {\bfseries 93} (2004) 152302},
  \href{https://arxiv.org/abs/nucl-ex/0401003}{{\ttfamily
  arXiv:nucl-ex/0401003}}.

\bibitem{STAR:2004qya}
{\bfseries STAR} Collaboration, J.~Adams {\em et~al.}, ``{Pion interferometry
  in Au+Au collisions at $\sqrt{s_{\mathrm{NN}}}$ = 200 GeV}'',
  \href{https://doi.org/10.1103/PhysRevC.71.044906}{{\em Phys. Rev. C}
  {\bfseries 71} (2005) 044906},
  \href{https://arxiv.org/abs/nucl-ex/0411036}{{\ttfamily
  arXiv:nucl-ex/0411036}}.

\bibitem{ALICE:2015hvw}
{\bfseries ALICE} Collaboration, J.~Adam {\em et~al.}, ``{One-dimensional pion,
  kaon, and proton femtoscopy in Pb-Pb collisions at $\sqrt{s_{\rm {NN}}}$
  =2.76 TeV}'', \href{https://doi.org/10.1103/PhysRevC.92.054908}{{\em Phys.
  Rev. C} {\bfseries 92} (2015) 054908},
  \href{https://arxiv.org/abs/1506.07884}{{\ttfamily arXiv:1506.07884
  [nucl-ex]}}.

\bibitem{ALICE:2020ibs}
{\bfseries ALICE} Collaboration, S.~Acharya {\em et~al.}, ``{Search for a
  common baryon source in high-multiplicity pp collisions at the LHC}'',
  \href{https://doi.org/10.1016/j.physletb.2020.135849}{{\em Phys. Lett. B}
  {\bfseries 811} (2020) 135849},
  \href{https://arxiv.org/abs/2004.08018}{{\ttfamily arXiv:2004.08018
  [nucl-ex]}}.

\bibitem{ALICE:2025wuy}
{\bfseries ALICE} Collaboration, S.~Acharya {\em et~al.}, ``{Femtoscopic study
  of the proton-proton and proton-deuteron systems in heavy-ion collisions at
  the LHC}'', \href{https://arxiv.org/abs/2505.01061}{{\ttfamily
  arXiv:2505.01061 [nucl-ex]}}.

\bibitem{Scheibl:1998tk}
R.~Scheibl and U.~W. Heinz, ``{Coalescence and flow in ultrarelativistic heavy
  ion collisions}'', \href{https://doi.org/10.1103/PhysRevC.59.1585}{{\em Phys.
  Rev. C} {\bfseries 59} (1999) 1585--1602},
  \href{https://arxiv.org/abs/nucl-th/9809092}{{\ttfamily
  arXiv:nucl-th/9809092}}.

\bibitem{Sun:2018mqq}
K.-J. Sun, C.~M. Ko, and B.~D\"onigus, ``{Suppression of light nuclei
  production in collisions of small systems at the Large Hadron Collider}'',
  \href{https://doi.org/10.1016/j.physletb.2019.03.033}{{\em Phys. Lett. B}
  {\bfseries 792} (2019) 132--137},
  \href{https://arxiv.org/abs/1812.05175}{{\ttfamily arXiv:1812.05175
  [nucl-th]}}.

\bibitem{Blum:2019suo}
K.~Blum and M.~Takimoto, ``{Nuclear coalescence from correlation functions}'',
  \href{https://doi.org/10.1103/PhysRevC.99.044913}{{\em Phys. Rev. C}
  {\bfseries 99} (2019) 044913},
  \href{https://arxiv.org/abs/1901.07088}{{\ttfamily arXiv:1901.07088
  [nucl-th]}}.

\bibitem{Bellini:2018epz}
F.~Bellini and A.~P. Kalweit, ``{Testing production scenarios for
  (anti-)(hyper-)nuclei and exotica at energies available at the CERN Large
  Hadron Collider}'', \href{https://doi.org/10.1103/PhysRevC.99.054905}{{\em
  Phys. Rev. C} {\bfseries 99} (2019) 054905},
  \href{https://arxiv.org/abs/1807.05894}{{\ttfamily arXiv:1807.05894
  [hep-ph]}}.

\bibitem{Bellini:2020cbj}
F.~Bellini, K.~Blum, A.~P. Kalweit, and M.~Puccio, ``{Examination of
  coalescence as the origin of nuclei in hadronic collisions}'',
  \href{https://doi.org/10.1103/PhysRevC.103.014907}{{\em Phys. Rev. C}
  {\bfseries 103} (2021) 014907},
  \href{https://arxiv.org/abs/2007.01750}{{\ttfamily arXiv:2007.01750
  [nucl-th]}}.

\bibitem{ALICE:2022xiu}
{\bfseries ALICE} Collaboration, S.~Acharya {\em et~al.}, ``{First Measurement
  of Antideuteron Number Fluctuations at Energies Available at the Large Hadron
  Collider}'', \href{https://doi.org/10.1103/PhysRevLett.131.041901}{{\em Phys.
  Rev. Lett.} {\bfseries 131} (2023) 041901},
  \href{https://arxiv.org/abs/2204.10166}{{\ttfamily arXiv:2204.10166
  [nucl-ex]}}.

\bibitem{Feckova:2016kjx}
Z.~Feckov\'a, J.~Steinheimer, B.~Tom\'a\v{s}ik, and M.~Bleicher, ``{Formation
  of deuterons by coalescence: Consequences for deuteron number
  fluctuations}'', \href{https://doi.org/10.1103/PhysRevC.93.054906}{{\em Phys.
  Rev. C} {\bfseries 93} (2016) 054906},
  \href{https://arxiv.org/abs/1603.05854}{{\ttfamily arXiv:1603.05854
  [nucl-th]}}.

\bibitem{Braun-Munzinger:2023gsd}
P.~Braun-Munzinger, K.~Redlich, A.~Rustamov, and J.~Stachel, ``{The imprint of
  conservation laws on correlated particle production}'',
  \href{https://doi.org/10.1007/JHEP08(2024)113}{{\em JHEP} {\bfseries 08}
  (2024) 113}, \href{https://arxiv.org/abs/2312.15534}{{\ttfamily
  arXiv:2312.15534 [nucl-th]}}.

\bibitem{Vovchenko:2020kwg}
V.~Vovchenko and V.~Koch, ``{Particlization of an interacting hadron resonance
  gas with global conservation laws for event-by-event fluctuations in
  heavy-ion collisions}'',
  \href{https://doi.org/10.1103/PhysRevC.103.044903}{{\em Phys. Rev. C}
  {\bfseries 103} (2021) 044903},
  \href{https://arxiv.org/abs/2012.09954}{{\ttfamily arXiv:2012.09954
  [hep-ph]}}.

\bibitem{Vovchenko:2019pjl}
V.~Vovchenko and H.~Stoecker, ``{Thermal-FIST: A package for heavy-ion
  collisions and hadronic equation of state}'',
  \href{https://doi.org/10.1016/j.cpc.2019.06.024}{{\em Comput. Phys. Commun.}
  {\bfseries 244} (2019) 295--310},
  \href{https://arxiv.org/abs/1901.05249}{{\ttfamily arXiv:1901.05249
  [nucl-th]}}.

\bibitem{ALICE:2022xpf}
{\bfseries ALICE} Collaboration, S.~Acharya {\em et~al.}, ``{Closing in on
  critical net-baryon fluctuations at LHC energies: Cumulants up to third order
  in Pb\textendash{}Pb collisions}'',
  \href{https://doi.org/10.1016/j.physletb.2022.137545}{{\em Phys. Lett. B}
  {\bfseries 844} (2023) 137545},
  \href{https://arxiv.org/abs/2206.03343}{{\ttfamily arXiv:2206.03343
  [nucl-ex]}}.

\bibitem{ALICE:2024rnr}
{\bfseries ALICE} Collaboration, S.~Acharya {\em et~al.}, ``{Probing
  Strangeness Hadronization with Event-by-Event Production of Multistrange
  Hadrons}'', \href{https://doi.org/10.1103/PhysRevLett.134.022303}{{\em Phys.
  Rev. Lett.} {\bfseries 134} (2025) 022303},
  \href{https://arxiv.org/abs/2405.19890}{{\ttfamily arXiv:2405.19890
  [nucl-ex]}}.

\bibitem{Aamodt:2008zz}
{\bfseries ALICE} Collaboration, K.~Aamodt {\em et~al.}, ``{The ALICE
  experiment at the CERN LHC}'',
  \href{https://doi.org/10.1088/1748-0221/3/08/S08002}{{\em JINST} {\bfseries
  3} (2008) S08002}.

\bibitem{Abelev:2014ffa}
{\bfseries ALICE} Collaboration, B.~B. Abelev {\em et~al.}, ``{Performance of
  the ALICE Experiment at the CERN LHC}'',
  \href{https://doi.org/10.1142/S0217751X14300440}{{\em Int. J. Mod. Phys. A}
  {\bfseries 29} (2014) 1430044},
  \href{https://arxiv.org/abs/1402.4476}{{\ttfamily arXiv:1402.4476
  [nucl-ex]}}.

\bibitem{Abbas:2013taa}
{\bfseries ALICE} Collaboration, E.~Abbas {\em et~al.}, ``{Performance of the
  ALICE VZERO system}'',
  \href{https://doi.org/10.1088/1748-0221/8/10/P10016}{{\em JINST} {\bfseries
  8} (2013) P10016}, \href{https://arxiv.org/abs/1306.3130}{{\ttfamily
  arXiv:1306.3130 [nucl-ex]}}.

\bibitem{Arslandok:2022dyb}
M.~Arslandok, E.~Hellb\"ar, M.~Ivanov, R.~H. M\"unzer, and J.~Wiechula,
  ``{Track Reconstruction in a High-Density Environment with ALICE}'',
  \href{https://doi.org/10.3390/particles5010008}{{\em Particles} {\bfseries 5}
  (2022) 84--95}, \href{https://arxiv.org/abs/2203.10325}{{\ttfamily
  arXiv:2203.10325 [physics.ins-det]}}.

\bibitem{Abelev:2013qoq}
{\bfseries ALICE} Collaboration, B.~Abelev {\em et~al.}, ``{Centrality
  determination of Pb-Pb collisions at $\sqrt{s_{\rm NN}}$ = 2.76 TeV with
  ALICE}'', \href{https://doi.org/10.1103/PhysRevC.88.044909}{{\em Phys. Rev.
  C} {\bfseries 88} (2013) 044909},
  \href{https://arxiv.org/abs/1301.4361}{{\ttfamily arXiv:1301.4361
  [nucl-ex]}}.

\bibitem{ALICE:2018tvk}
{\bfseries ALICE} Collaboration, S.~Acharya {\em et~al.}, ``{Centrality
  determination in heavy ion collisions}'', {\em ALICE-PUBLIC-2018-011} (2018)
  . \url{https://cds.cern.ch/record/2636623}.

\bibitem{Alme:2010ke}
J.~Alme {\em et~al.}, ``{The ALICE TPC, a large 3-dimensional tracking device
  with fast readout for ultra-high multiplicity events}'',
  \href{https://doi.org/10.1016/j.nima.2010.04.042}{{\em Nucl. Instrum. Meth.
  A} {\bfseries 622} (2010) 316--367},
  \href{https://arxiv.org/abs/1001.1950}{{\ttfamily arXiv:1001.1950
  [physics.ins-det]}}.

\bibitem{ALICE:2013xmt}
{\bfseries ALICE} Collaboration, B.~B. Abelev {\em et~al.}, ``{Multi-strange
  baryon production at mid-rapidity in Pb-Pb collisions at $\sqrt{s_{\rm NN}}$
  = 2.76 TeV}'', \href{https://doi.org/10.1016/j.physletb.2014.05.052}{{\em
  Phys. Lett. B} {\bfseries 728} (2014) 216--227},
  \href{https://arxiv.org/abs/1307.5543}{{\ttfamily arXiv:1307.5543
  [nucl-ex]}}. [Erratum: Phys.Lett.B 734, 409--410 (2014)].

\bibitem{ALICE:2013wgn}
{\bfseries ALICE} Collaboration, B.~B. Abelev {\em et~al.}, ``{Multiplicity
  Dependence of Pion, Kaon, Proton and Lambda Production in p-Pb Collisions at
  $\sqrt{s_{\rm NN}}$ = 5.02 TeV}'',
  \href{https://doi.org/10.1016/j.physletb.2013.11.020}{{\em Phys. Lett. B}
  {\bfseries 728} (2014) 25--38},
  \href{https://arxiv.org/abs/1307.6796}{{\ttfamily arXiv:1307.6796
  [nucl-ex]}}.

\bibitem{ALICE:2016fzo}
{\bfseries ALICE} Collaboration, J.~Adam {\em et~al.}, ``{Enhanced production
  of multi-strange hadrons in high-multiplicity proton-proton collisions}'',
  \href{https://doi.org/10.1038/nphys4111}{{\em Nature Phys.} {\bfseries 13}
  (2017) 535--539}, \href{https://arxiv.org/abs/1606.07424}{{\ttfamily
  arXiv:1606.07424 [nucl-ex]}}.

\bibitem{Akindinov:2013tea}
A.~Akindinov {\em et~al.}, ``{Performance of the ALICE Time-Of-Flight detector
  at the LHC}'', \href{https://doi.org/10.1140/epjp/i2013-13044-x}{{\em Eur.
  Phys. J. Plus} {\bfseries 128} (2013) 44}.

\bibitem{ParticleDataGroup:2024cfk}
{\bfseries Particle Data Group} Collaboration, S.~Navas {\em et~al.}, ``{Review
  of particle physics}'',
  \href{https://doi.org/10.1103/PhysRevD.110.030001}{{\em Phys. Rev. D}
  {\bfseries 110} (2024) 030001}.

\bibitem{Wang:1991hta}
X.-N. Wang and M.~Gyulassy, ``{HIJING: A Monte Carlo model for multiple jet
  production in p p, p A and A A collisions}'',
  \href{https://doi.org/10.1103/PhysRevD.44.3501}{{\em Phys. Rev. D} {\bfseries
  44} (1991) 3501--3516}.

\bibitem{GEANT4:2002zbu}
{\bfseries $\textsc{geant4}$} Collaboration, S.~Agostinelli {\em et~al.},
  ``{$\textsc{geant4}$--a simulation toolkit}'',
  \href{https://doi.org/10.1016/S0168-9002(03)01368-8}{{\em Nucl. Instrum.
  Meth. A} {\bfseries 506} (2003) 250--303}.

\bibitem{Nonaka:2017kko}
T.~Nonaka, M.~Kitazawa, and S.~Esumi, ``{More efficient formulas for efficiency
  correction of cumulants and effect of using averaged efficiency}'',
  \href{https://doi.org/10.1103/PhysRevC.95.064912}{{\em Phys. Rev. C}
  {\bfseries 95} (2017) 064912},
  \href{https://arxiv.org/abs/1702.07106}{{\ttfamily arXiv:1702.07106
  [physics.data-an]}}. [Erratum: Phys.Rev.C 103, 029901 (2021)].

\bibitem{Braun-Munzinger:2016yjz}
P.~Braun-Munzinger, A.~Rustamov, and J.~Stachel, ``{Bridging the gap between
  event-by-event fluctuation measurements and theory predictions in
  relativistic nuclear collisions}'',
  \href{https://doi.org/10.1016/j.nuclphysa.2017.01.011}{{\em Nucl. Phys. A}
  {\bfseries 960} (2017) 114--130},
  \href{https://arxiv.org/abs/1612.00702}{{\ttfamily arXiv:1612.00702
  [nucl-th]}}.

\bibitem{ALICE:2023ulv}
{\bfseries ALICE} Collaboration, S.~Acharya {\em et~al.}, ``{Measurements of
  Chemical Potentials in Pb-Pb Collisions at $\sqrt{s_{\rm NN}} = 5.02$ TeV}'',
  \href{https://doi.org/10.1103/PhysRevLett.133.092301}{{\em Phys. Rev. Lett.}
  {\bfseries 133} (2024) 092301},
  \href{https://arxiv.org/abs/2311.13332}{{\ttfamily arXiv:2311.13332
  [nucl-ex]}}.

\bibitem{Rustamov:2022sqm}
A.~Rustamov, J.~Stroth, and R.~Holzmann, ``{A model-free procedure to correct
  for volume fluctuations in E-by-E analyses of particle multiplicities}'',
  \href{https://doi.org/10.1016/j.nuclphysa.2023.122641}{{\em Nucl. Phys. A}
  {\bfseries 1034} (2023) 122641},
  \href{https://arxiv.org/abs/2211.14849}{{\ttfamily arXiv:2211.14849
  [nucl-th]}}.

\bibitem{Holzmann:2024wyd}
R.~Holzmann, V.~Koch, A.~Rustamov, and J.~Stroth, ``{Controlling volume
  fluctuations for studies of critical phenomena in nuclear collisions}'',
  \href{https://doi.org/10.1016/j.nuclphysa.2024.122924}{{\em Nucl. Phys. A}
  {\bfseries 1050} (2024) 122924},
  \href{https://arxiv.org/abs/2403.03598}{{\ttfamily arXiv:2403.03598
  [nucl-th]}}.

\bibitem{Anticic:2013htn}
T.~Anticic {\em et~al.}, ``{Phase-space dependence of particle-ratio
  fluctuations in Pb + Pb collisions from 20 A to 158 A GeV beam energy}'',
  \href{https://doi.org/10.1103/PhysRevC.89.054902}{{\em Phys. Rev. C}
  {\bfseries 89} (2014) 054902},
  \href{https://arxiv.org/abs/1310.3428}{{\ttfamily arXiv:1310.3428
  [nucl-ex]}}.

\bibitem{ALICE:2017jsh}
{\bfseries ALICE} Collaboration, S.~Acharya {\em et~al.}, ``{Relative particle
  yield fluctuations in $\text{ Pb-Pb }$ collisions at $\sqrt{s_\mathrm{{NN}}}
  =2.76\hbox { TeV}$}'',
  \href{https://doi.org/10.1140/epjc/s10052-019-6711-x}{{\em Eur. Phys. J. C}
  {\bfseries 79} (2019) 236},
  \href{https://arxiv.org/abs/1712.07929}{{\ttfamily arXiv:1712.07929
  [nucl-ex]}}.

\bibitem{Pandav:2018bdx}
A.~Pandav, D.~Mallick, and B.~Mohanty, ``{Effect of limited statistics on
  higher order cumulants measurement in heavy-ion collision experiments}'',
  \href{https://doi.org/10.1016/j.nuclphysa.2019.08.002}{{\em Nucl. Phys. A}
  {\bfseries 991} (2019) 121608},
  \href{https://arxiv.org/abs/1809.08892}{{\ttfamily arXiv:1809.08892
  [nucl-ex]}}.

\bibitem{Vovchenko:2019kes}
V.~Vovchenko, B.~D\"onigus, and H.~Stoecker, ``{Canonical statistical model
  analysis of p-p , p -Pb, and Pb-Pb collisions at energies available at the
  CERN Large Hadron Collider}'',
  \href{https://doi.org/10.1103/PhysRevC.100.054906}{{\em Phys. Rev. C}
  {\bfseries 100} (2019) 054906},
  \href{https://arxiv.org/abs/1906.03145}{{\ttfamily arXiv:1906.03145
  [hep-ph]}}.

\bibitem{ALICE:2019hno}
{\bfseries ALICE} Collaboration, S.~Acharya {\em et~al.}, ``{Production of
  charged pions, kaons, and (anti-)protons in Pb-Pb and inelastic $pp$
  collisions at $\sqrt{s_{\rm NN}}$ = 5.02 TeV}'',
  \href{https://doi.org/10.1103/PhysRevC.101.044907}{{\em Phys. Rev. C}
  {\bfseries 101} (2020) 044907},
  \href{https://arxiv.org/abs/1910.07678}{{\ttfamily arXiv:1910.07678
  [nucl-ex]}}.

\bibitem{ALICE:2020nkc}
{\bfseries ALICE} Collaboration, S.~Acharya {\em et~al.}, ``{Multiplicity
  dependence of $\pi $, K, and p production in pp collisions at $\sqrt{s} = 13$
  TeV}'', \href{https://doi.org/10.1140/epjc/s10052-020-8125-1}{{\em Eur. Phys.
  J. C} {\bfseries 80} (2020) 693},
  \href{https://arxiv.org/abs/2003.02394}{{\ttfamily arXiv:2003.02394
  [nucl-ex]}}.

\bibitem{STAR:2020ddh}
{\bfseries STAR} Collaboration, J.~Adam {\em et~al.}, ``{Beam energy dependence
  of net- {\ensuremath{\Lambda}} fluctuations measured by the STAR experiment
  at the BNL Relativistic Heavy Ion Collider}'',
  \href{https://doi.org/10.1103/PhysRevC.102.024903}{{\em Phys. Rev. C}
  {\bfseries 102} (2020) 024903},
  \href{https://arxiv.org/abs/2001.06419}{{\ttfamily arXiv:2001.06419
  [nucl-ex]}}.

\bibitem{ALICE:2025cqy}
{\bfseries ALICE} Collaboration, I.~J. Abualrob {\em et~al.}, ``{Centrality
  dependence of strange particle production in Pb-Pb collisions at
  $\sqrt{s_{\rm NN}} = 5.02$ TeV}'',
  \href{https://doi.org/10.1103/j21y-pqd6}{{\em Phys. Rev. C} {\bfseries 114}
  (2026) 024909}, \href{https://arxiv.org/abs/2511.10360}{{\ttfamily
  arXiv:2511.10360 [nucl-ex]}}.

\bibitem{Mahlein:2023fmx}
M.~Mahlein, L.~Barioglio, F.~Bellini, L.~Fabbietti, C.~Pinto, B.~Singh, and
  S.~Tripathy, ``{A realistic coalescence model for deuteron production}'',
  \href{https://doi.org/10.1140/epjc/s10052-023-11972-3}{{\em Eur. Phys. J. C}
  {\bfseries 83} (2023) 804},
  \href{https://arxiv.org/abs/2302.12696}{{\ttfamily arXiv:2302.12696
  [hep-ex]}}.

\bibitem{Mahlein:2024pur}
M.~Mahlein, C.~Pinto, and L.~Fabbietti, ``{ToMCCA: a Toy Monte Carlo
  Coalescence Afterburner}'',
  \href{https://doi.org/10.1140/epjc/s10052-024-13486-y}{{\em Eur. Phys. J. C}
  {\bfseries 84} (2024) 1136},
  \href{https://arxiv.org/abs/2404.03352}{{\ttfamily arXiv:2404.03352
  [hep-ph]}}.

\bibitem{Wiringa:1994wb}
R.~B. Wiringa, V.~G.~J. Stoks, and R.~Schiavilla, ``{An Accurate
  nucleon-nucleon potential with charge independence breaking}'',
  \href{https://doi.org/10.1103/PhysRevC.51.38}{{\em Phys. Rev. C} {\bfseries
  51} (1995) 38--51}, \href{https://arxiv.org/abs/nucl-th/9408016}{{\ttfamily
  arXiv:nucl-th/9408016}}.

\bibitem{Vovchenko:2024pvk}
V.~Vovchenko, ``{Density correlations under global and local charge
  conservation}'', \href{https://doi.org/10.1103/PhysRevC.110.L061902}{{\em
  Phys. Rev. C} {\bfseries 110} (2024) L061902},
  \href{https://arxiv.org/abs/2409.01397}{{\ttfamily arXiv:2409.01397
  [hep-ph]}}.

\end{thebibliography}\endgroup

\newpage
\appendix
\makeatletter
\renewcommand{\thesubsection}{\arabic{subsection}}
\makeatother

\section*{End Matter}
\renewcommand{\thefigure}{\arabic{figure}}
\renewcommand{\theequation}{\arabic{equation}}
\setcounter{figure}{0}
\setcounter{equation}{0}

\subsection*{Cumulants of $\Lambda$, $\bar{\Lambda}$, and net-$\Lambda$}

\begin{figure}[h]
    \centering
    \includegraphics[width=0.7\textwidth]{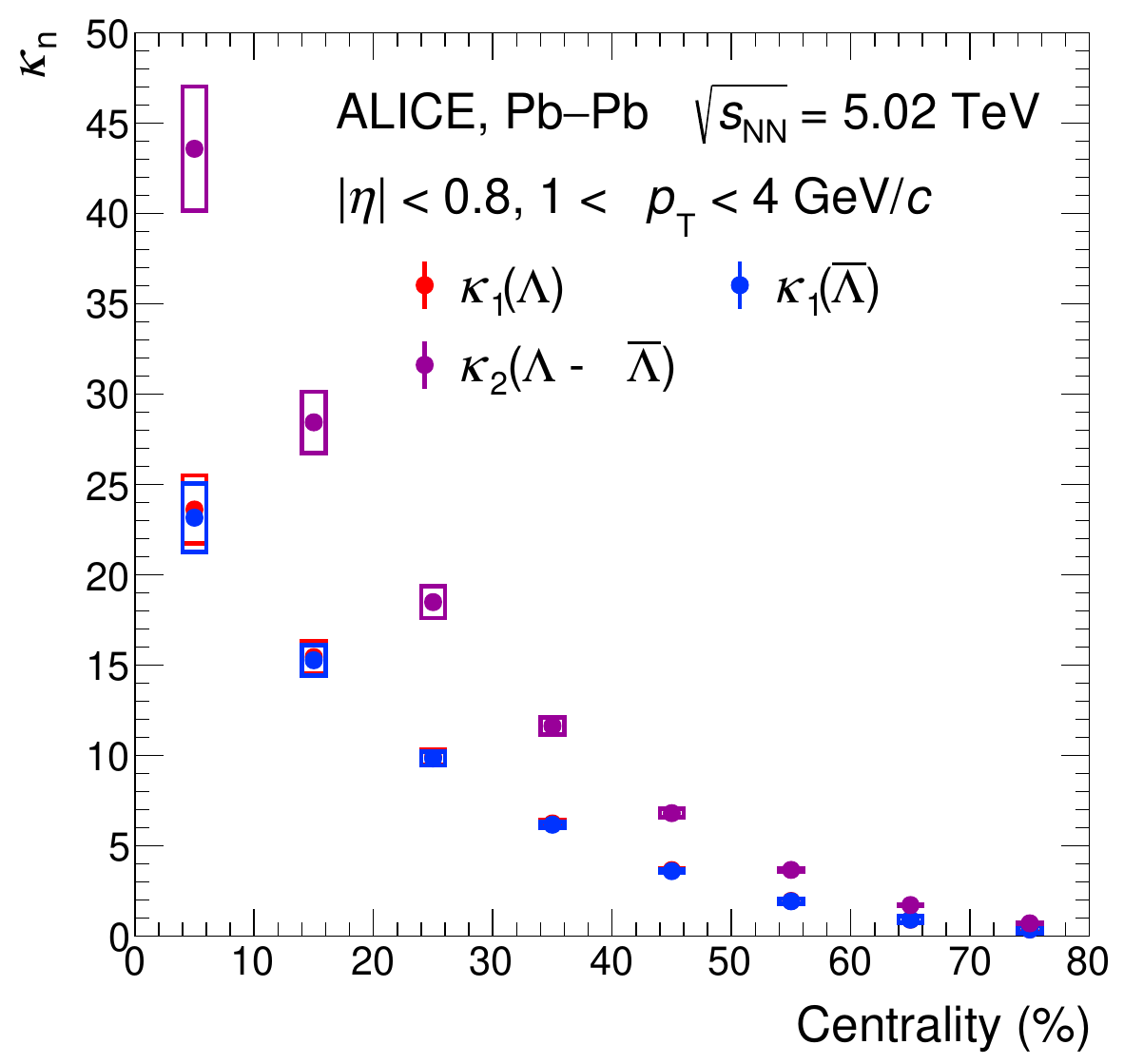}
    \caption{First order cumulants of the $\Lambda$ number and $\bar{\Lambda}$ number and second order cumulant of the net-$\Lambda$ number in Pb--Pb collisions at \snn = 5.02 TeV, as a function of centrality.}
    \label{fig:cumul}
\end{figure}

In \cref{fig:cumul}, the first order cumulants of $\Lambda$ and $\bar{\Lambda}$, $\kappa_{1}(\Lambda)$ and $\kappa_{1}(\bar\Lambda)$, and the second order cumulant of net-$\Lambda$, $\kappa_{2}(\Lambda - \bar{\Lambda})$, are shown as a function of centrality. The results are corrected by the reconstruction and selection efficiency computed from MC simulations.
As the first order cumulants and $\kappa_{2}(\Lambda - \bar\Lambda)$ are not affected by volume fluctuations, no additional correction is applied in these cases.
Statistical and systematic uncertainties are computed using the same approaches described in the main text.

\clearpage

\section{The ALICE Collaboration}
\label{app:collab}
\begin{flushleft} 
\small

D.A.H.~Abdallah\,\orcidlink{0000-0003-4768-2718}\,$^{\rm 133}$, 
M.M.~Abedin$^{\rm 4}$, 
I.J.~Abualrob\,\orcidlink{0009-0005-3519-5631}\,$^{\rm 110}$, 
S.~Acharya\,\orcidlink{0000-0002-9213-5329}\,$^{\rm 49}$, 
A.B.~Adiguzel$^{\rm 90}$, 
K.~Agarwal\,\orcidlink{0000-0001-5781-3393}\,$^{\rm II,}$$^{\rm 23}$, 
G.~Aglieri Rinella\,\orcidlink{0000-0002-9611-3696}\,$^{\rm 32}$, 
L.~Aglietta\,\orcidlink{0009-0003-0763-6802}\,$^{\rm 24}$, 
N.~Agrawal\,\orcidlink{0000-0003-0348-9836}\,$^{\rm 25}$, 
Z.~Ahammed\,\orcidlink{0000-0001-5241-7412}\,$^{\rm 131}$
% ...

\section*{Affiliation Notes}

$^{\rm I}$ Deceased\\
$^{\rm II}$ Also at: INFN Trieste, Trieste, Italy\\
$^{\rm III}$ Also at: Fondazione Bruno Kessler (FBK), Trento, Italy\\
$^{\rm IV}$ Also at: Czech Technical University in Prague, Prague, Czech Republic\\
$^{\rm V}$ Also at: Dipartimento DET del Politecnico di Torino, Turin, Italy\\
$^{\rm VI}$ Also at: University College of Dublin, Dublin, Ireland\\
$^{\rm VII}$ Also at: Department of Applied Physics, Aligarh Muslim University, Aligarh, India\\
$^{\rm VIII}$ Also at: Institute of Theoretical Physics, University of Wroclaw, Wroclaw, Poland\\
$^{\rm IX}$ Also at: Facultad de Ciencias, Universidad Nacional Aut\'{o}noma de M\'{e}xico, Mexico City, Mexico\\
$^{\rm X}$ Also at: Lovely Professional University, Punjab, India
\\

\section*{Collaboration Institutes}

$^{1}$ A.I. Alikhanyan National Science Laboratory (Yerevan Physics Institute) Foundation, Yerevan, Armenia\\
$^{2}$ AGH University of Krakow, Cracow, Poland\\
$^{3}$ Bogolyubov Institute for Theoretical Physics, National Academy of Sciences of Ukraine, Kyiv, Ukraine\\
$^{4}$ Bose Institute, Department of Physics  and Centre for Astroparticle Physics and Space Science (CAPSS), Kolkata, India\\
$^{5}$ California Polytechnic State University, San Luis Obispo, California, United States\\
$^{6}$ Central China Normal University, Wuhan, China\\
$^{7}$ Centro de Aplicaciones Tecnol\'{o}gicas y Desarrollo Nuclear (CEADEN), Havana, Cuba\\
$^{8}$ Centro de Investigaci\'{o}n y de Estudios Avanzados (CINVESTAV), Mexico City and M\'{e}rida, Mexico\\
$^{9}$ Chicago State University, Chicago, Illinois, United States\\
$^{10}$ China Nuclear Data Center, China Institute of Atomic Energy, Beijing, China\\
$^{11}$ China University of Geosciences, Wuhan, China\\
$^{12}$ Chungbuk National University, Cheongju, Republic of Korea\\
$^{13}$ Comenius University Bratislava, Faculty of Mathematics, Physics and Informatics, Bratislava, Slovak Republic\\
$^{14}$ Creighton University, Omaha, Nebraska, United States\\
$^{15}$ Department of Physics, Aligarh Muslim University, Aligarh, India\\
$^{16}$ Department of Physics, Pusan National University, Pusan, Republic of Korea\\
$^{17}$ Department of Physics, Sejong University, Seoul, Republic of Korea\\
$^{18}$ Department of Physics, University of California, Berkeley, California, United States\\
$^{19}$ Department of Physics, University of Oslo, Oslo, Norway\\
$^{20}$ Department of Physics and Technology, University of Bergen, Bergen, Norway\\
$^{21}$ Dipartimento di Fisica, Universit\`{a} di Pavia, Pavia, Italy\\
$^{22}$ Dipartimento di Fisica dell'Universit\`{a} and Sezione INFN, Cagliari, Italy\\
$^{23}$ Dipartimento di Fisica dell'Universit\`{a} and Sezione INFN, Trieste, Italy\\
$^{24}$ Dipartimento di Fisica dell'Universit\`{a} and Sezione INFN, Turin, Italy\\
$^{25}$ Dipartimento di Fisica e Astronomia dell'Universit\`{a} and Sezione INFN, Bologna, Italy\\
$^{26}$ Dipartimento di Fisica e Astronomia dell'Universit\`{a} and Sezione INFN, Catania, Italy\\
$^{27}$ Dipartimento di Fisica e Astronomia dell'Universit\`{a} and Sezione INFN, Padova, Italy\\
$^{28}$ Dipartimento di Fisica `E.R.~Caianiello' dell'Universit\`{a} and Gruppo Collegato INFN, Salerno, Italy\\
$^{29}$ Dipartimento DISAT del Politecnico and Sezione INFN, Turin, Italy\\
$^{30}$ Dipartimento di Scienze MIFT, Universit\`{a} di Messina, Messina, Italy\\
$^{31}$ Dipartimento Interateneo di Fisica `M.~Merlin' and Sezione INFN, Bari, Italy\\
$^{32}$ European Organization for Nuclear Research (CERN), Geneva, Switzerland\\
$^{33}$ Faculty of Electrical Engineering, Mechanical Engineering and Naval Architecture, University of Split, Split, Croatia\\
$^{34}$ Faculty of Nuclear Sciences and Physical Engineering, Czech Technical University in Prague, Prague, Czech Republic\\
$^{35}$ Faculty of Physics, Sofia University, Sofia, Bulgaria\\
$^{36}$ Faculty of Science, P.J.~\v{S}af\'{a}rik University, Ko\v{s}ice, Slovak Republic\\
$^{37}$ Faculty of Technology, Environmental and Social Sciences, Bergen, Norway\\
$^{38}$ Frankfurt Institute for Advanced Studies, Johann Wolfgang Goethe-Universit\"{a}t Frankfurt, Frankfurt, Germany\\
$^{39}$ Fudan University, Shanghai, China\\
$^{40}$ Gauhati University, Department of Physics, Guwahati, India\\
$^{41}$ Helmholtz-Institut f\"{u}r Strahlen- und Kernphysik, Rheinische Friedrich-Wilhelms-Universit\"{a}t Bonn, Bonn, Germany\\
$^{42}$ Helsinki Institute of Physics (HIP), Helsinki, Finland\\
$^{43}$ High Energy Physics Group,  Universidad Aut\'{o}noma de Puebla, Puebla, Mexico\\
$^{44}$ Horia Hulubei National Institute of Physics and Nuclear Engineering, Bucharest, Romania\\
$^{45}$ HUN-REN Wigner Research Centre for Physics, Budapest, Hungary\\
$^{46}$ Indian Institute of Technology Bombay (IIT), Mumbai, India\\
$^{47}$ Indian Institute of Technology Indore, Indore, India\\
$^{48}$ INFN, Laboratori Nazionali di Frascati, Frascati, Italy\\
$^{49}$ INFN, Sezione di Bari, Bari, Italy\\
$^{50}$ INFN, Sezione di Bologna, Bologna, Italy\\
$^{51}$ INFN, Sezione di Cagliari, Cagliari, Italy\\
$^{52}$ INFN, Sezione di Catania, Catania, Italy\\
$^{53}$ INFN, Sezione di Padova, Padova, Italy\\
$^{54}$ INFN, Sezione di Pavia, Pavia, Italy\\
$^{55}$ INFN, Sezione di Torino, Turin, Italy\\
$^{56}$ INFN, Sezione di Trieste, Trieste, Italy\\
$^{57}$ Inha University, Incheon, Republic of Korea\\
$^{58}$ Institute for Gravitational and Subatomic Physics (GRASP), Utrecht University/Nikhef, Utrecht, Netherlands\\
$^{59}$ Institute of Experimental Physics, Slovak Academy of Sciences, Ko\v{s}ice, Slovak Republic\\
$^{60}$ Institute of Physics of the Czech Academy of Sciences, Prague, Czech Republic\\
$^{61}$ Institute of Space Science (ISS), Bucharest, Romania\\
$^{62}$ Institut f\"{u}r Kernphysik, Johann Wolfgang Goethe-Universit\"{a}t Frankfurt, Frankfurt, Germany\\
$^{63}$ Instituto de Ciencias Nucleares, Universidad Nacional Aut\'{o}noma de M\'{e}xico, Mexico City, Mexico\\
$^{64}$ Instituto de F\'{i}sica, Universidade Federal do Rio Grande do Sul (UFRGS), Porto Alegre, Brazil\\
$^{65}$ Instituto de F\'{\i}sica, Universidad Nacional Aut\'{o}noma de M\'{e}xico, Mexico City, Mexico\\
$^{66}$ iThemba LABS, National Research Foundation, Somerset West, South Africa\\
$^{67}$ Jeonbuk National University, Jeonju, Republic of Korea\\
$^{68}$ Korea Institute of Science and Technology Information, Daejeon, Republic of Korea\\
$^{69}$ Laboratoire de Physique Subatomique et de Cosmologie, Universit\'{e} Grenoble-Alpes, CNRS-IN2P3, Grenoble, France\\
$^{70}$ Lawrence Berkeley National Laboratory, Berkeley, California, United States\\
$^{71}$ Lund University Department of Physics, Division of Particle Physics, Lund, Sweden\\
$^{72}$ Marietta Blau Institute, Vienna, Austria\\
$^{73}$ Nagasaki Institute of Applied Science, Nagasaki, Japan\\
$^{74}$ Nara Women{'}s University (NWU), Nara, Japan\\
$^{75}$ National Centre for Nuclear Research, Warsaw, Poland\\
$^{76}$ National Institute of Science Education and Research, Homi Bhabha National Institute, Jatni, India\\
$^{77}$ National Nuclear Research Center, Baku, Azerbaijan\\
$^{78}$ National Research and Innovation Agency - BRIN, Jakarta, Indonesia\\
$^{79}$ Niels Bohr Institute, University of Copenhagen, Copenhagen, Denmark\\
$^{80}$ Nikhef, National institute for subatomic physics, Amsterdam, Netherlands\\
$^{81}$ Nuclear Physics Group, STFC Daresbury Laboratory, Daresbury, United Kingdom\\
$^{82}$ Nuclear Physics Institute of the Czech Academy of Sciences, Husinec-\v{R}e\v{z}, Czech Republic\\
$^{83}$ Oak Ridge National Laboratory, Oak Ridge, Tennessee, United States\\
$^{84}$ Ohio State University, Columbus, Ohio, United States\\
$^{85}$ Physics Department, Panjab University, Chandigarh, India\\
$^{86}$ Physics Department, University of Jammu, Jammu, India\\
$^{87}$ Physics Program and International Institute for Sustainability with Knotted Chiral Meta Matter (WPI-SKCM$^{2}$), Hiroshima University, Hiroshima, Japan\\
$^{88}$ Physikalisches Institut, Eberhard-Karls-Universit\"{a}t T\"{u}bingen, T\"{u}bingen, Germany\\
$^{89}$ Physikalisches Institut, Ruprecht-Karls-Universit\"{a}t Heidelberg, Heidelberg, Germany\\
$^{90}$ Physik Department, Technische Universit\"{a}t M\"{u}nchen, Munich, Germany\\
$^{91}$ Politecnico di Bari and Sezione INFN, Bari, Italy\\
$^{92}$ Research Division and ExtreMe Matter Institute EMMI, GSI Helmholtzzentrum f\"ur Schwerionenforschung GmbH, Darmstadt, Germany\\
$^{93}$ Saga University, Saga, Japan\\
$^{94}$ Saha Institute of Nuclear Physics, Homi Bhabha National Institute, Kolkata, India\\
$^{95}$ School of Physics and Astronomy, University of Birmingham, Birmingham, United Kingdom\\
$^{96}$ Secci\'{o}n F\'{\i}sica, Departamento de Ciencias, Pontificia Universidad Cat\'{o}lica del Per\'{u}, Lima, Peru\\
$^{97}$ SUBATECH, IMT Atlantique, Nantes Universit\'{e}, CNRS-IN2P3, Nantes, France\\
$^{98}$ Sungkyunkwan University, Suwon City, Republic of Korea\\
$^{99}$ Suranaree University of Technology, Nakhon Ratchasima, Thailand\\
$^{100}$ Technical University of Ko\v{s}ice, Ko\v{s}ice, Slovak Republic\\
$^{101}$ The Henryk Niewodniczanski Institute of Nuclear Physics, Polish Academy of Sciences, Cracow, Poland\\
$^{102}$ The University of Texas at Austin, Austin, Texas, United States\\
$^{103}$ Universidad Aut\'{o}noma de Sinaloa, Culiac\'{a}n, Mexico\\
$^{104}$ Universidade de S\~{a}o Paulo (USP), S\~{a}o Paulo, Brazil\\
$^{105}$ Universidade Estadual de Campinas (UNICAMP), Campinas, Brazil\\
$^{106}$ Universidade Federal do ABC, Santo Andre, Brazil\\
$^{107}$ Universitatea Nationala de Stiinta si Tehnologie Politehnica Bucuresti, Bucharest, Romania\\
$^{108}$ University of Cape Town, Cape Town, South Africa\\
$^{109}$ University of Derby, Derby, United Kingdom\\
$^{110}$ University of Houston, Houston, Texas, United States\\
$^{111}$ University of Jyv\"{a}skyl\"{a}, Jyv\"{a}skyl\"{a}, Finland\\
$^{112}$ University of Kansas, Lawrence, Kansas, United States\\
$^{113}$ University of Liverpool, Liverpool, United Kingdom\\
$^{114}$ University of Science and Technology of China, Hefei, China\\
$^{115}$ University of Silesia in Katowice, Katowice, Poland\\
$^{116}$ University of South-Eastern Norway, Kongsberg, Norway\\
$^{117}$ University of Tennessee, Knoxville, Tennessee, United States\\
$^{118}$ University of the Witwatersrand, Johannesburg, South Africa\\
$^{119}$ University of Tokyo, Tokyo, Japan\\
$^{120}$ University of Tsukuba, Tsukuba, Japan\\
$^{121}$ University of Zagreb Faculty of Science, Department of Physics, Zagreb, Croatia\\
$^{122}$ Universit\"{a}t M\"{u}nster, Institut f\"{u}r Kernphysik, M\"{u}nster, Germany\\
$^{123}$ Universit\'{e} Clermont Auvergne, CNRS/IN2P3, LPC, Clermont-Ferrand, France\\
$^{124}$ Universit\'{e} de Lyon, CNRS/IN2P3, Institut de Physique des 2 Infinis de Lyon, Lyon, France\\
$^{125}$ Universit\'{e} de Strasbourg, CNRS, IPHC UMR 7178, F-67000 Strasbourg, France, Strasbourg, France\\
$^{126}$ Universit\'{e} Paris-Saclay, Centre d'Etudes de Saclay (CEA), IRFU, D\'{e}partment de Physique Nucl\'{e}aire (DPhN), Saclay, France\\
$^{127}$ Universit\'{e}  Paris-Saclay, CNRS/IN2P3, IJCLab, Orsay, France\\
$^{128}$ Universit\`{a} degli Studi di Foggia, Foggia, Italy\\
$^{129}$ Universit\`{a} del Piemonte Orientale, Vercelli, Italy\\
$^{130}$ Universit\`{a} di Brescia, Brescia, Italy\\
$^{131}$ Variable Energy Cyclotron Centre, Homi Bhabha National Institute, Kolkata, India\\
$^{132}$ Warsaw University of Technology, Warsaw, Poland\\
$^{133}$ Wayne State University, Detroit, Michigan, United States\\
$^{134}$ Yale University, New Haven, Connecticut, United States\\
$^{135}$ Yildiz Technical University, Istanbul, Turkey\\
$^{136}$ Yonsei University, Seoul, Republic of Korea\\
$^{137}$ Affiliated with an institute formerly covered by a cooperation agreement with CERN\\
$^{138}$ Affiliated with an international laboratory covered by a cooperation agreement with CERN.\\

\end{flushleft} 
\end{document}